\documentclass[aps,prl,showkeys,showpacs,twocolumn,longbibliography,superscriptaddress,notitlepage,floatfix]{revtex4-2}

\usepackage{amsmath,amsfonts,amssymb,color,epsfig,graphics,graphicx,latexsym,theorem,url,multirow,diagbox}
\usepackage[colorlinks,colorlinks,citecolor=blue,linkcolor=blue,urlcolor=blue]{hyperref}
\usepackage[utf8]{inputenc}
\usepackage{booktabs}
\usepackage[T1]{fontenc}
\usepackage{courier}
\usepackage{listings}
\usepackage{dcolumn}
\usepackage{comment}
\usepackage{times}

\usepackage{algorithm}
\usepackage{algorithmic} 
\usepackage{braket}
\usepackage{bm}

\begin{document}

\title{Quantum federated learning through blind quantum computing}

\author{Weikang Li}\thanks{These authors contributed equally to this work.}
\affiliation{Center for Quantum Information, IIIS, Tsinghua University, Beijing
100084, People\textquoteright s Republic of China}

\author{Sirui Lu}\thanks{These authors contributed equally to this work.}
\affiliation{Center for Quantum Information, IIIS, Tsinghua University, Beijing
100084, People\textquoteright s Republic of China}
\affiliation{Max-Planck-Institut f\"ur Quantenoptik, Hans-Kopfermann-Str.\ 1, D-85748 Garching, Germany}

\author{Dong-Ling Deng}
\email{dldeng@tsinghua.edu.cn}
\affiliation{Center for Quantum Information, IIIS, Tsinghua University, Beijing
100084, People\textquoteright s Republic of China}
\affiliation{Shanghai Qi Zhi Institute, 41th Floor, AI Tower, No. 701 Yunjin Road, Xuhui District, Shanghai 200232, China}

\begin{abstract}
Private distributed learning studies the problem of how multiple distributed entities collaboratively train a shared deep network with their private data unrevealed. 
With the security provided by the protocols of blind quantum computation,
the cooperation between quantum physics and machine learning may lead to unparalleled prospect for solving private distributed learning tasks.
In this paper,
we introduce a quantum protocol for distributed learning that is able to utilize the computational power of the remote quantum servers while keeping the private data safe.
For concreteness, we first introduce a protocol for private single-party delegated training of variational quantum classifiers based on blind quantum computing and then extend this protocol to multiparty private distributed learning incorporated with differential privacy. We carry out extensive numerical simulations with different real-life datasets and encoding strategies to benchmark the effectiveness  of our protocol.  We find that our protocol is robust to experimental imperfections and is secure under the gradient attack after the incorporation of differential privacy. Our results show the potential for handling computationally expensive distributed learning tasks with privacy guarantees, 
thus providing a valuable guide for exploring quantum advantages from the security perspective in the field of machine learning with real-life applications.

\end{abstract}



\maketitle

\begin{figure}
\includegraphics[width=0.46\textwidth]{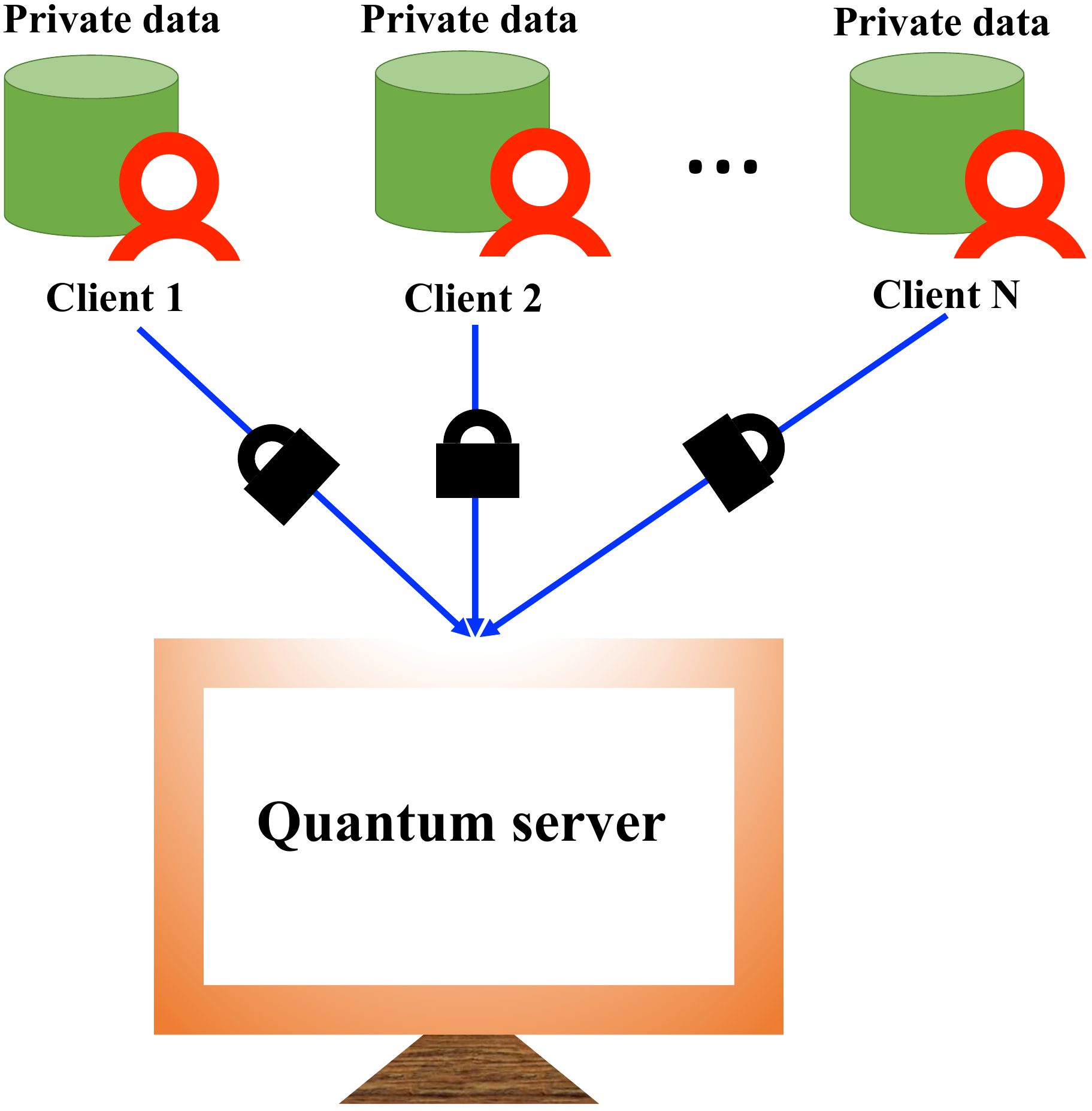}
\caption{Schematic illustration of private distributed learning with a quantum server, where multiple clients can utilize the computational power from the quantum server to collaboratively train a shared model while leaking no information about their private data during the computing process.}
\label{Schematic illustration}
\end{figure}

Quantum machine learning is an emergent interdisciplinary field that explores the interplay between machine learning and quantum physics \cite{Biamonte2017Quantum,Dunjko2018Machine,Sarma2019Machine}. Recently, it has attracted tremendous attention across different communities.  On the one hand, machine learning, or more broadly artificial intelligence \cite{Russell2020Artificial}, has achieved momentous success over the past decade \cite{Lecun2015Deep,Jordan2015Machine} and a number of notoriously challenging problems, such as, playing the game of Go \cite{Silver2016Mastering, Silver2017Mastering} or predicting protein structures \cite{Senior2020Improved}, have been cracked successfully. This brings new intriguing opportunities for utilizing machine learning to tackle outstanding problems in quantum science \cite{Dunjko2018Machine,Sarma2019Machine,Carleo2019Machine}.
On the other hand, as manifested by the experimental demonstration of quantum supremacy \cite{Arute2019Quantum,Zhong2020Quantum}, 
the field of quantum computing \cite{Nielsen2010Quantum} has also made remarkable progress in recent years, giving rise to new striking possibilities of enhancing machine learning with quantum devices, in turn. Along this direction,  a variety of works have shown strong evidence that
quantum computers could outperform classical computers in solving certain  machine learning problems \cite{Biamonte2017Quantum,Dunjko2018Machine,Sarma2019Machine}. Notable examples  include the Harrow-Hassidim-Lloyd algorithm \cite{Harrow2009Quantum}, quantum principal component analysis \cite{Lloyd2014Quantum}, quantum-enhanced feature space \cite{Havlicek2019Supervised,Schuld2019Quantum},  quantum generative models \cite{Gao2018Quantum,Lloyd2018Quantum,Hu2019Quantum},  quantum support vector machines \cite{Rebentrost2014Quantum}, etc.  Most of these existing works study the possible quantum advantages from the perspective of quantum speedup in machine learning.  In this paper, we  explore possible quantum advantages from the security perspective and introduce a quantum protocol for private distributed learning based on blind quantum computing \cite{Barz2012Demonstration,Broadbent2009Universal,Childs2001Secure,Fitzsimons2017Unconditionally,
Fitzsimons2017Private,Giovannetti2013Efficient,Morimae2013Secure,Sheng2015Deterministic} (see Fig.~\ref{Schematic illustration} for a pictorial illustration).

Private distributed learning studies the interplay between privacy and machine learning, which is of crucial importance nowadays in the data-intensive and sensitive society
\cite{Zerka2020Systematic}.
Indeed, in many collaborative learning tasks \cite{Hosny2018Artificial,Chen2019Gmail},  it is highly desirable that the privacy for each party is preserved. 
To this end, a popular proposal developed in recent years concerns federated learning
\cite{Konecny2016Federated,McMahan2017Communication},
where given a shared machine learning model,
the updates of the model rather than the sensitive data are uploaded to the public server.
In addition, to better fight against the malicious attacks that try to recover the sensitive data from the updated parameters of the model (such as the gradients),
the differential privacy method that adds noise to the shared updates has been extensively studied over the years \cite{Dwork2008Differential,Dwork2014Algorithmic,Abadi2016Deep}. More recently, another approach to private distributed learning has also been introduced (dubbed InstaHide), which encrypts each training image with a "one-time secret key" consisting of mixing several randomly chosen images and applying a random pixel-wise mask \cite{Huang2020Instahide}. However, each of these approaches bears its pros and cons, and exhibits different levels of security. For instance, the differential privacy method often comes at the cost of accuracy  and 
does not apply to side-channel computations performed by malicious parties \cite{Bassily2014Private,Beimel2014Bounds}. Fully-homomorphic encryption \cite{Gentry2009Fully} can guarantee   privacy against arbitrary side-computations by adversary during training. Yet, it is impractical in modern deep learning settings due to the high computational overheads and requirement of special setups.  In the quantum domain, Sheng and Zhou have proposed a distributed secure quantum machine learning protocol recently,  where  a classical client with limited quantum power can delegate a learning task to a quantum server with the private data preserved \cite{Sheng2017Distributed}.

 In this paper,
we introduce a protocol for private distributed learning, including federated learning in particular, based on  blind quantum computing \cite{Barz2012Demonstration,Broadbent2009Universal,Childs2001Secure,Fitzsimons2017Unconditionally,
Fitzsimons2017Private,Giovannetti2013Efficient,Morimae2013Secure,Sheng2015Deterministic}.
We first introduce a protocol for private single-party delegated training with variational quantum classifiers \cite{Cerezo2020Variational}, and then extend this protocol to multiparty distributed learning as shown in Fig.~\ref{Schematic illustration},
where there are several parties that wish to collaboratively train a shared model, 
e.g., 
several hospitals with inadequate and sensitive patient data on each side want to train a shared model for disease diagnosis. To better protect the privacy for each client, we incorporate the differential privacy method into the protocol. We carry out extensive numerical simulations with different real-life datasets (such as the MNIST dataset about handwritten digit images  and the Wisconsin Diagnostic Breast Cancer dataset) and encoding strategies to benchmark the effectiveness  of our protocol. In addition, we also demonstrate that our protocol is robust to experimental imperfections and is secure under the renowned  gradient attack after the incorporation of differential privacy. We stress two intriguing advantages of our quantum private distributed learning protocol compared with its classical counterparts, such as federated learning. First, in our protocol the clients can both utilize the computational power from the quantum server and use differential privacy to protect the private information without trusting the server, while in classical settings the clients need either enough local computational power or a trustworthy server. Second, certain quantum learning algorithms that may exhibit exponential advantages, such as the quantum generative model \cite{Gao2018Quantum} or the quantum kernel estimation for classification problems \cite{Liu2020Rigorous}, can also be adapted to our protocol. As a result, our protocol has the potential to achieve exponential advantages for these learning tasks. Our results open a new avenue for exploring quantum advantages in tackling practical machine learning tasks, which would be crucial for achieving large-scale unconditionally secure private distributed learning with future quantum technologies.


\begin{figure}
\includegraphics[width=0.485\textwidth]{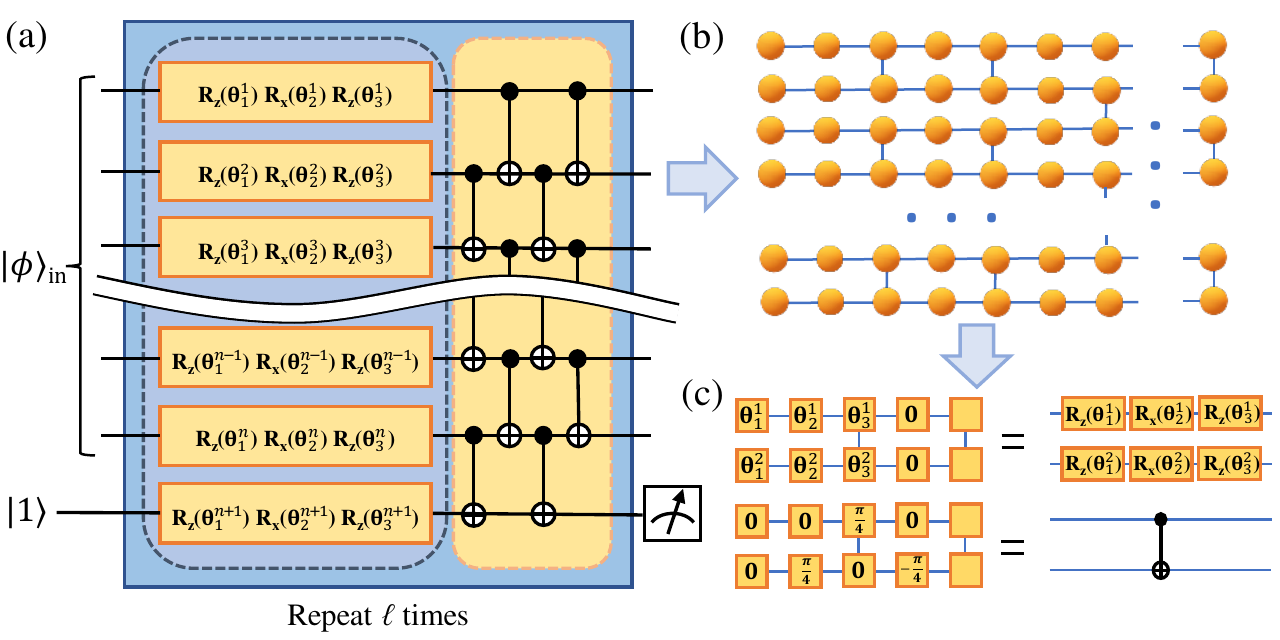}
\caption{(a) The structure of the variational quantum classifiers that conveniently fits the protocol of universal blind quantum computation (UBQC) in \cite{Broadbent2009Universal}. This basic structure contains a rotation layer with z- and x- rotations  and an entangling layer with controlled-NOT gates, and it will repeat $\ell$ times to form the desired classifier.   (b) The brickwork state used in the UBQC protocol. (c) The implementations of the rotations and controlled-NOT gates by measuring the brickwork state along specific directions. See \cite{PQDLSupp} for details.}
\label{VQC}
\end{figure}

\emph{The general recipe.}\textemdash
To start with, we first recap the protocol of universal blind quantum computation (UBQC) introduced in 
\cite{Broadbent2009Universal}, which allows a client Alice to delegate her computation task to a quantum server Bob without revealing any information about the  inputs, outputs and computation to Bob. In this protocol, we assume that Alice is able to prepare single qubits randomly chosen from a finite set and send them to the server. But, other than this she  does not need to have any other quantum computational power or quantum memory. Bob has the balance of the required quantum computational resources but may not be trusted by Alice. After Bob received the qubits prepared by Alice, they proceed with a two-way classical communication that allows Alice to drive the computation by giving single-qubit measurement instructions (which depend on previous measurement outcomes) to Bob. At the end, Alice can deduce the desired computational results but Bob learns nothing about Alice's inputs, outputs, and the computation. This protocol is unconditional secure, in the sense that it does not rely on any computational assumptions and holds regardless of what actions a cheating Bob undertakes.


For our purpose, we take advantage of the security and integrity provided by the UBQC protocol and consider delegating quantum machine learning tasks to an untrusted quantum server. For simplicity, we focus on the classification tasks and consider a scenario where $N$ clients want to train collaboratively a shared quantum classifier with their private data unrevealed.  To this end, each client can delegate the computation to the server by using the UBQC protocol \cite{Broadbent2009Universal}. 
For the training process,
we define the loss function as the cross entropy 
\begin{eqnarray}
L\left(h\left(|\psi\rangle_{\mathrm{in}} ; \Theta\right), \mathbf{a}\right)=-\sum_{k} a_{k} \log g_{k},
\end{eqnarray}
where  $\mathbf{a}\equiv (a_1, \cdots, a_m)$  is the label corresponding to the input state $\ket{\phi}_{\text{in}}$ in the form of one-hot encoding, $h$ denotes  the hypothesis function determined by the variational quantum classifier (with parameters collectively denoted by $ \Theta$), and $\mathbf{g}\equiv (g_1,\cdots, g_m)=\text{diag} (\rho_{\text{out}})$ gives all the probabilities of the corresponding classification categories with $\rho_{\text{out}}$ denoting the output state of the classifier \cite{Lu2020Quantum}.   
To minimize the loss function, 
the "parameter shift rule"  \cite{Mitarai2018Quantum,Li2017Hybrid,Schuld2019Evaluating} can be used as a subroutine
to compute the gradients (see the Supplementary Material \cite{PQDLSupp} for details).
When $N=1$, we only have a single client and we can assume that the quantum classifier is kept in confidential. In this case, the client Alice can use the UBQC protocol directly and her privacy is preserved perfectly. Indeed, the server even does not know that Alice is dealing with a machine learning task. Whereas, when $N$ is larger than one  multiple clients share the same quantum classifier. In this case, it is more reasonable to assume that the quantum classifier and its update information are publicly announced. As a result, an adversary may exploit these information to retrieve the clients' private data, through for instance the gradient attack method \cite{Zhu2019Deep}.  To better protect the privacy for each client, we should incorporate the differential privacy method to our protocol, as discussed below.

\begin{figure}
\includegraphics[width=0.48\textwidth]{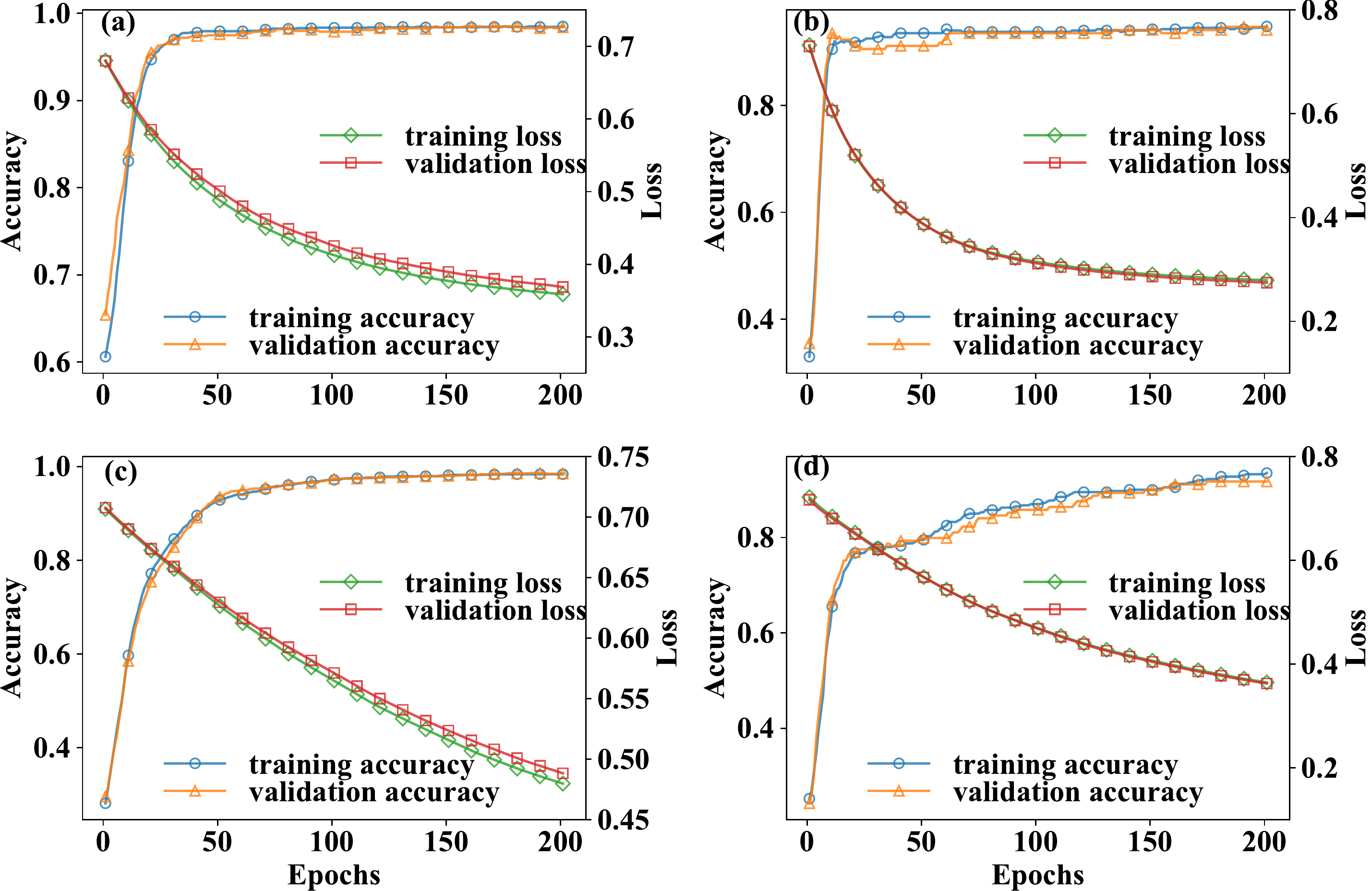}
\caption{Numerical results on the demonstration of private quantum distributed learning based on the universal blind quantum computing protocol introduced in \cite{Broadbent2009Universal}. (a) and (b) show the results for the single-client case with the MNIST dataset and the Wisconsin Diagnostic Breast Cancer dataset, respectively. Similarly, (c) and (d) show the results for the multi-client case for the same two datasets, with differential privacy adapted. See \cite{PQDLSupp} for details.
}
\label{VQC_train}
\end{figure}

\emph{Single-party delegated learning.}\textemdash
We first consider the $N=1$ case. Suppose Alice has a variational quantum classifier with structures shown in Fig.~\ref{VQC} (a). This classifier can be decomposed into rotation and entangling layers with rotation and the controlled-NOT gates that are conveniently implemented by measuring each site of the brickwork state along different angles, as illustrated in Fig.~\ref{VQC} (b,c). Alice wants to finish a classification task with this quantum classifier, but she does not have a quantum computer and thus needs to delegate the  task to Bob who maintains a quantum server. As mentioned in the above discussion, she can accomplish this through applying the UBQC protocol directly. For concreteness, here we consider two classification tasks, one about identifying handwritten digit images "1 or 9" in the MNIST dataset
\cite{LeCun1998Mnist}  and the other about classifying data samples in the Wisconsin Diagnostic Breast Cancer (WDBC) dataset
\cite{H.Wolberg1992Uci}. We numerically simulate the training and validation process \cite{PQDLSupp} and our results are plotted in Fig.~\ref{VQC_train} (a, b). From these figures, we find that both the training and validation accuracies increase rapidly at the beginning of the training process and then saturate roughly at $98\%$ and $94\%$ for the MNIST and WDBC datasets, respectively. This shows  that the delegated learning indeed can achieve a decent performance on these classification tasks. 
 In addition, considering the unavoidable imperfections  of quantum devices in reality, the preparation of the single qubits and the transfer of these qubits  on the Alice's side and the measurement on the Bob's side will inevitably bring some noises. To benchmark the robustness of this protocol, we  also carry out numerical simulations with various noises and different noise strengths. Our detailed results are shown in the Supplementary Materials \cite{PQDLSupp}.

We remark that we deliberately choose  the variational quantum classifier above because it is "native" to measurement-based quantum computing and can be conveniently implemented when applying the UBQC protocol. One may choose other variational quantum classifiers, such as the hierarchical quantum classifiers   \cite{Grant2018Hierarchical}  or quantum convolutional neural networks \cite{Cong2019Quantum}. Yet, for these classifiers some additional modifications might be needed so as to match the particular brickwork state used in the UBQC protocol. In addition, we stress that  in this single-client scenario Alice can keep the structure of her quantum classifier confidential, and Bob learns nothing about Alice's inputs, outputs, or desired computation. Alice's privacy can be perfectly preserved.

\begin{figure}
\begin{algorithm}[H]
\caption{Quantum private distributed learning}
\label{Algo_PDL}
\begin{algorithmic}
\REQUIRE The untrained model $h$ with parameters $\Theta$, the loss function $L$, the number of clients $N$, the number of samples belongs to one client $n$, the batch size $n_b$, the number of iterations $T$, the learning rate $\epsilon$, the Adam optimizer $f_{\text{Adam}}$, the gradient norm bound $R$, the Laplace noise $\mathcal{N}$, the noise strength $\mu$
\ENSURE The trained model
\STATE Initialization: randomly generate a length-$T$ string $S$ with each element corresponding to an index of a client
\FOR{ $i \in [T]$}
    \STATE Choose the client with index $S_i$ and randomly choose $n_b$ samples among the $n$ samples of the client
    \STATE Calculate the gradient under the UBQC protocol $\vec{g}_i\leftarrow\frac{1}{n_b}\Sigma_{k=1}^{n_b}\nabla L(h(\ket{\psi}_{\text{i,k}};\Theta),\mathbf{a}_{\text{i,k}})$
    \STATE Clip the gradient: for every element $\vec{g}_i[j]$ in $\vec{g}_i$, $\overline{g}_i[j]\leftarrow\vec{g}_i[j]/\text{max}(1,\frac{||\vec{g}_i[j]||}{R})$
    \STATE Add noise: for every element $\overline{g}_i[j]$ in $\overline{g}_i$, $\widetilde{g}_i[j]\leftarrow\overline{g}_i[j]+\frac{2R}{\mu}\mathcal{N}(0, 1)$
    \STATE Updates: $\Theta\leftarrow\Theta-\epsilon \cdot f_{\text{Adam}}(\widetilde{g}_i)$
\ENDFOR
\STATE Output the trained model
\end{algorithmic}
\end{algorithm}
\end{figure}

\begin{figure}
\begin{algorithm}[H]
\caption{Gradient attack algorithm}
\label{Algo_GA}
\begin{algorithmic}
\REQUIRE The target gradient $\vec{g}_{t}$, the model $h$ with parameters $\Theta$, the loss function $L$ for training the model, the string $S$ that contains $n$ different labels, the number of iterations $T$, the learning rate $\epsilon$, the Adam optimizer $f_{\text{Adam}}$
\ENSURE The recovered input data and label
\STATE Prepare the initial state as a uniform quantum state $\ket{\phi}\leftarrow\ket{\phi_0}$
\FOR{$\mathbf{k} \in S$}
    \STATE Define the loss function for the gradient attack as $L_{\text{GA}}(\ket{\phi}) \leftarrow ||\nabla L(h(\ket{\phi};\Theta),\mathbf{k})-\vec{g}_{t}||^2$
    \FOR{$i \in [T]$}
        \STATE Calculate the gradient $\ket{\vec{g}_i}\leftarrow\nabla_{\text{in}} L_{\text{GA}}(\ket{\phi})$
        \STATE Updates: $\ket{\phi}\leftarrow\ket{\phi}-\epsilon \cdot f_{\text{Adam}}(\ket{\vec{g}_i})$
        \STATE Normalization: $\ket{\phi}\leftarrow\frac{1}{||\ket{\phi}||}\ket{\phi}$
    \ENDFOR
    \STATE If the final loss function with this label $\mathbf{k}$ is the minimum, the final state $\ket{\phi}$ and label $\mathbf{k}$ will be set as the recovered input data and label
\ENDFOR
\STATE Output the recovered input data and label
\end{algorithmic}
\end{algorithm}
\end{figure}

\emph{Multi-client private distributed learning.}\textemdash
We now turn to the case of $N>1$, where several clients collaboratively train a shared model with their private data unrevealed. 
One may regard this as the quantum counterpart of classical federated learning \cite{Konecny2016Federated,McMahan2017Communication}.
Assume that each client has $n$ training samples and the shared quantum classifier has the same structure as that in Fig.~\ref{VQC} (a). In every turn of the collaborative training,
one client will use the UBQC protocol to delegate the computation of the gradient to the server with the private data the client holds. Then the gradients will be uploaded to update  the parameters of the shared quantum classifier with the following rule:
\begin{eqnarray}
\Theta_{t+1} = \Theta_{t} - \epsilon \cdot f(\vec{g}),
\end{eqnarray}
where $\Theta_{t} $ denotes collectively the parameters at the $t$-th step, $\epsilon$ is the learning rate, $\vec{g}$ is the vectorized gradient, and $f(\vec{g})$ is a modification function of the gradient which we take the Adam optimizer
\cite{Kingma2014Adam} 
in this paper.
However, 
if there are eavesdroppers with access to the model,
uploading the gradients directly can be attacked by reverse engineering to recover the input corresponding to the gradients 
\cite{Zhu2019Deep}.
To handle this problem,
we adapt the idea of differential privacy \cite{Dwork2008Differential,Dwork2014Algorithmic,Abadi2016Deep} to our protocol,  where the clients who obtain the desired gradients can add appropriate noise such as Laplace noise or Gaussian noise to these gradients before uploading them.
The feasibility can be understood as follows.
Before adding noise,
there might be an input whose gradient exactly corresponds to the uploaded gradient,
thus the real input can be recovered by some optimization process.
With adaption of differential privacy,
the attacker only sees a combination of the real gradient and random noise.
The result of the attack becomes random and it is hard for the attacker to differentiate the gradient corresponding to the recovered input and the real gradient obtained from the original data.
 In our work, the process of adding noise has two steps. After obtaining the gradients, the clients firstly clip every element of the gradient vector to have a fixed bound $R$. Then the normally distributed noises with mean value zero and standard deviation $\frac{2R}{\mu}$ will be added to every element of the gradient vector independently, where $\mu$ can be used to control the noise strengths.
Yet, it is also worthwhile to point out that in this case the clients are training a shared model which is public. So if the updated parameters from a client can make the model behave better, it is inevitable that some information about this client's data will be leaked. This is in sharp contrast to the single-client case, where the privacy can be perfect as discussed above.  In real experiments, the intrinsic quantum noises may offer the randomness the protocol needs and be beneficial for privacy protection in these tasks
\cite{Du2020Quantuma}.
With this strategy,
the clients can iteratively train the shared model to find the optimal parameters.
The pseudocode for our protocol is given in Algorithm \ref{Algo_PDL}. 
To benchmark how our protocol works, we also carry out numerical simulations on the MNIST and WDBC datasets. Our results are shown in Fig.~\ref{VQC_train}(c) and Fig.~\ref{VQC_train}(d), from which it is clear that the increase of the accuracies are slower than that for the single-client case. This is due to the adaption of 
differential privacy, which will downgrade the accuracy in general. To account for experimental imperfections, we also carry out numerical simulations with different noise strengths and our results are shown in the Supplementary Materials \cite{PQDLSupp}.

%

\begin{figure}
\includegraphics[width=0.48\textwidth]{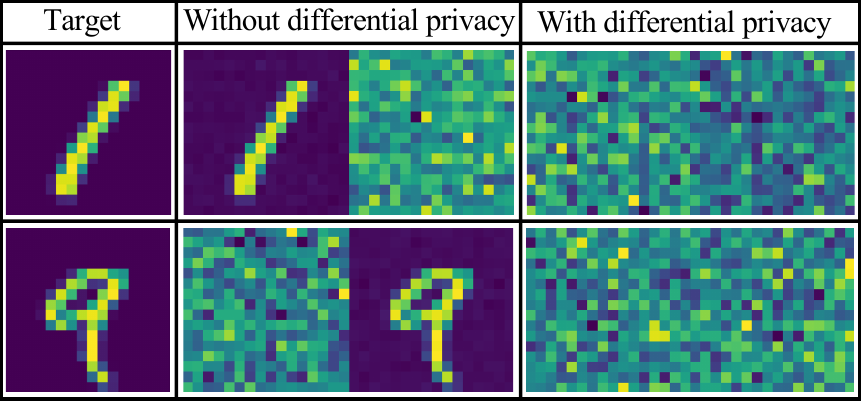}
\caption{The performance of the gradient attack when trying to recover the input data based on the updates the client uploaded. Here, we simulate the attacks numerically with two handwritten digit images of  "1" and "9". For each image, the attacker obtains two recovered images, rather than one image since he does not know the true label and needs to try both labels,   through this gradient attack method.  Without differential privacy, one of the recovered images has very high fidelity with the original image, thus the private data information is leaked.  Whereas with differential privacy, both the recovered images seem random and the private  data of the client is indeed preserved. 
}
\label{GA_train}
\end{figure}

\emph{Gradient attack.}\textemdash
So far we have presented a concrete protocol for private distributed learning with the method of differential privacy to prevent malicious attacks. To better illustrate why differential privacy is needed, we now apply the gradient attack \cite{Zhu2019Deep} 
to test the performance before and after adding noise to the gradients. Generally, when training a variational quantum circuit to minimize a certain loss function, we will calculate the derivative of this loss function with respect to the model parameters.
However, if an eavesdropper wants to attack a client's data, what he has is the updated parameters.
So inversely, the task becomes recovering the input of the circuit with the updates.
We first demonstrate how the gradient attack works in the situation where we upload the gradients directly without adding noises. We take the two handwritten digits with labels "1" and "9" respectively as a client's input to be attacked and suppose the attack happens at step $t$ with model parameters $\Theta$.
When the client uses the real input $\ket{\phi_{r}}$ to do the training,
the gradient descent method will give the target gradient $\vec{g}_{t} = \nabla L(h(\ket{\phi_{r}};\Theta),\mathbf{a})$.
On the other hand,
since the eavesdropper does not know any information about the input data as promised by the UBQC protocol,
he may starts with a uniform state $\ket{\phi}=\ket{\phi_0}$.
Now the loss function for gradient attack is defined as 
\begin{eqnarray}
L_{\text{GA}}(\ket{\phi}) = ||\nabla L(h(\ket{\phi};\Theta),\mathbf{a})-\nabla L(h(\ket{\phi_{r}};\Theta),\mathbf{a})||^2, \nonumber
\end{eqnarray}
which measures the distance between the current gradient to the target.
We mention that the eavesdropper has no information about the label $\mathbf{a}$ which is contained in the loss function,
so the attack needs to be carried out with all the possible labels.
To minimize this loss function,
we need to compute the derivatives of it with respect to the input samples rather than the circuit parameters,
and the concrete algorithm is given in Algorithm \ref{Algo_GA}.
The numerical demonstration for this algorithm to attack the handwritten digits without differential privacy is shown in the second column of Fig.~\ref{GA_train}, 
where the input digits are recovered with very high fidelity.
However,
if we clip the gradients with a bound $R$ and add appropriate Laplace noise to the uploaded gradients which is taken by Algorithm \ref{Algo_PDL} as a subroutine and will slightly reduce the training accuracy,
the loss function becomes
\begin{eqnarray}
L_{\text{GA}}(\ket{\phi}) = ||\nabla L(h(\ket{\phi};\Theta),\mathbf{a})-\mathcal{N}(\vec{g}_{t})||^2,
\end{eqnarray}
where $\mathcal{N}$ contains the operations of adding noise and clipping the gradients.
It turns out that the recovered digits from the noisy gradients have a clear difference from the real ones as shown in the third column of Fig.~\ref{GA_train}. It should be noted that even though the noise can help with the privacy in the figures,
the attacker may also get some information about the labels of the input according to the curve of the training process,
i.e.,
the loss function with the right label may achieve the smallest value.
One way to strengthen the privacy is to compute the gradients with several samples with different labels at a step, e.g. stochastic gradient descent with the size of each batch larger than one, and take their average value as the updates.

\emph{Discussion and conclusion.}\textemdash 
Although we focus on classification problems in this paper, our protocol carries over straightforwardly to other quantum learning problems as well. We remark that, the blind quantum computing scheme adapted in our protocol is universal: any computation that can be efficiently accomplished by a quantum circuit can be translated into the blind quantum computing scheme efficiently \cite{Broadbent2009Universal}. As a result, a quantum-circuit based learning algorithm that may achieve exponential advantages (e.g., the quantum generative model \cite{Gao2018Quantum} and the quantum kernel estimation for classification problems \cite{Liu2020Rigorous})  can also be translated straightforwardly to the blind quantum computing scenario, hence extended to our protocol as well. Consequently, for these tasks our protocol holds the potential to achieve exponential advantages.  
For instance, in Ref.~\cite{Gao2018Quantum} a quantum machine learning algorithm based on generative models has been introduced and rigorously proved to have exponential speedups over any classical algorithms for some instances if a quantum computer cannot be efficiently simulated classically. Now, considering training the quantum generative model in the setting of federated learning. By using our protocol, each client can delegate the computing of the gradients, which may cost exponential time on a classical computer, to the quantum server by using the UBQC subroutine. In this way, our protocol may achieve both exponential speedup and privacy protection.
In addition, our protocol requires that the clients have the capability of preparing single qubits, which may not be satisfied under certain circumstances. To release this requirement, one could exploit  other blind quantum computing protocols concerning purely classical clients \cite{Broadbent2009Universal,Huang2017Experimental,Mantri2017Flow}. Given the fact that privacy protection becomes more and more important in the data-intensive and sensitive society, an experimental demonstration of quantum private distributed learning would be a crucial step toward practical applications of quantum technologies for tackling security related problems in artificial intelligence in the future.

In summary, we have introduced a protocol for quantum private distributed learning based on blind quantum computing, where  several clients with inadequate data can collaboratively train a quantum learning model while keeping their sensitive data unrevealed. To benchmark the effectiveness of our protocol, we carried out extensive numerical simulations with different real-life datasets and encoding strategies, taking into consideration  experimental imperfections.  In addition, we also demonstrated that our protocol survives the well-known gradient attack  after the adaption of differential privacy.  Our results show the intriguing potential of achieving large-scale  private distributed learning with both near-term and future quantum technologies, and provide a valuable guide for exploring quantum advantages in real-life machine learning applications  from the security perspective.

We thank Haoyu Guo, Wenjie Jiang, Zhide Lu, Peixin Shen for helpful discussions. 
This work is supported by the start-up fund from Tsinghua University (Grant. No. 53330300320), 
the National Natural Science Foundation of China (Grant. No. 12075128), 
and the Shanghai Qi Zhi Institute.

\bibliography{QMLBib}

\clearpage

\setcounter{secnumdepth}{3}

\makeatletter
\setcounter{figure}{0}
\setcounter{equation}{0}
\renewcommand{\thefigure}{S\@arabic\c@figure}
\renewcommand \theequation{S\@arabic\c@equation}
\renewcommand \thetable{S\@arabic\c@table}
\renewcommand \thealgorithm{S\@arabic\c@algorithm}

\begin{center} 
	{\large \bf Supplementary Material: Quantum federated learning through blind quantum computing}
\end{center} 

In this Supplementary Material, we first give a brief review of the UBQC protocol \cite{Broadbent2009Universal}
and illustrate how to adapt our learning tasks to the UBQC protocol.
Then we will give the detailed settings of our numerical experiments in the following.
As mentioned in the main text,
we will  test the robustness of the quantum classifiers with different noise strengths. 
More numerical simulations of differential privacy with different noise strengths will be presented as well.
In the last part,
we will test a different data encoding strategy that encodes the data in the rotation angles through the quantum circuit and make some discussions about it.

\section{Review and usage of the universal blind quantum computation}

\begin{figure}[H]
\includegraphics[width=0.48\textwidth]{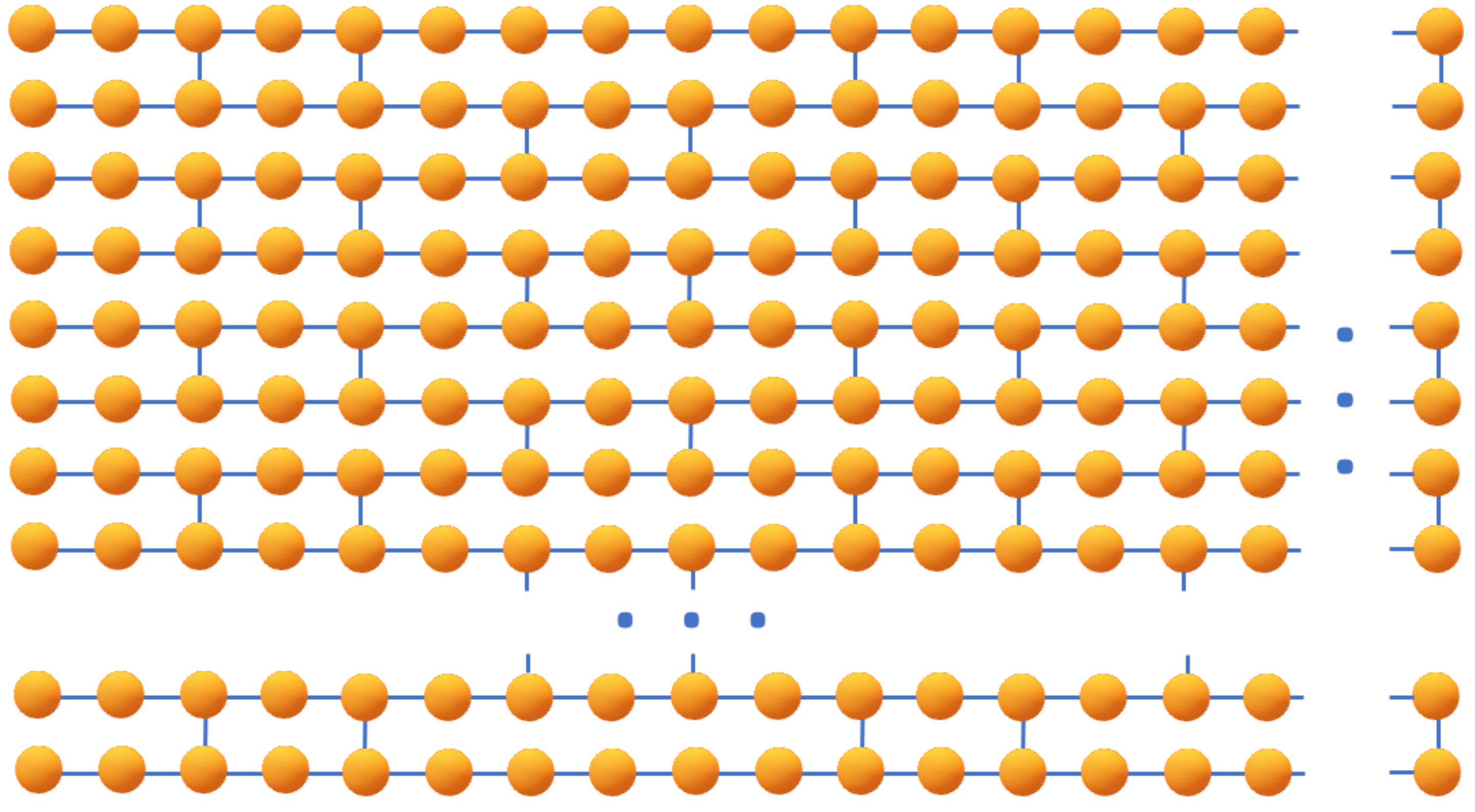}
\caption{The structure for a universal blind quantum computation scheme on the server's side adapted from \cite{Broadbent2009Universal}. Every orange ball denotes a qubit received from Alice, and the connected lines denote controlled-Z gates.}
\label{UBQC}
\end{figure}

The goal of blind quantum computation is to delegate a computational task to a quantum server while keeping the inputs, outputs, and computation details private.
Suppose there is a client Alice and a quantum server Bob.
The universal blind quantum computation from 
\cite{Broadbent2009Universal} 
can be achieved where Alice only needs the quantum power to prepare single qubits and sends them to Bob.
To start with,
the structure on the Bob's side which is called the brickwork state is shown in Fig.~\ref{UBQC}.
For each qubit in this structure,
Alice prepares a state in $\frac{1}{\sqrt{2}}(\ket{0}+e^{i \theta}\ket{1})$ where $\theta$ is a uniformly random number from a finite or infinite set and sends it to the corresponding spot.
After all the qubit preparations, 
Bob will apply controlled-Z gates to all the pairs connected by lines to create an entangled state.
Then the final state is suitable for measurement based quantum computations and Alice can delegate the computation by two-way classical communications.

Suppose there are $n$ columns and $m$ rows in the brickwork structure.
Let $\phi_{x,y}$ be the measurement angle on column x, row y to achieve the computation given the condition that the initial sent qubits are all in state $\frac{1}{\sqrt{2}}(\ket{0}+\ket{1})$.
Alice will have these angles in mind and give the measurement angle $\delta_{x, y}=\phi_{x, y}+\theta_{x, y}+\pi r_{x, y}$ to Bob where $\theta_{x, y}$ is the random angle for point $(x,y)$ in the single qubit preparations and $r_{x, y}$ is a uniformly random binary number.
It can be proved that in this way, 
correctness and blindness can both be achieved.
For the correctness,
the measurement on the state $\frac{1}{\sqrt{2}}(\ket{0}+e^{i (\phi+\theta)}\ket{1})$ in the basis $\frac{1}{\sqrt{2}}(\ket{0}\pm e^{i (\psi+\theta)}\ket{1})$ is the same as the measurement on the state $\frac{1}{\sqrt{2}}(\ket{0}+e^{i \phi}\ket{1})$ in the basis $\frac{1}{\sqrt{2}}(\ket{0}\pm e^{i \psi}\ket{1})$,
thus leaving no influence on the original computation.
As for the random binary number $r_{x, y}$, if it is zero, the measurement angle is not affected.
If it is one, Alice can simply flip the measurement result to get the correct answer.
For the blindness,
the information that Bob has can be divided into three parts:
(1) the single qubits sent from Alice;
(2) the classical information from Alice to apply the measurement based computation; 
(3) the binary information from the measurement outcomes.
Firstly,
the uniformly random number $\theta_{x, y}$ can make the information encoded in the qubit and the classical information sent by Alice uniformly random on Bob's side.
Secondly,
the binary measurement outcomes will appear uniformly random on Bob's side as promised by $r_{x, y}$.
Furthermore,
an authentication protocol to detect a cheating server has also been proposed in 
\cite{Broadbent2009Universal} in details.
Thus the feasibility of this protocol has a theoretical guarantee.

\begin{figure}[H]
\includegraphics[width=0.48\textwidth]{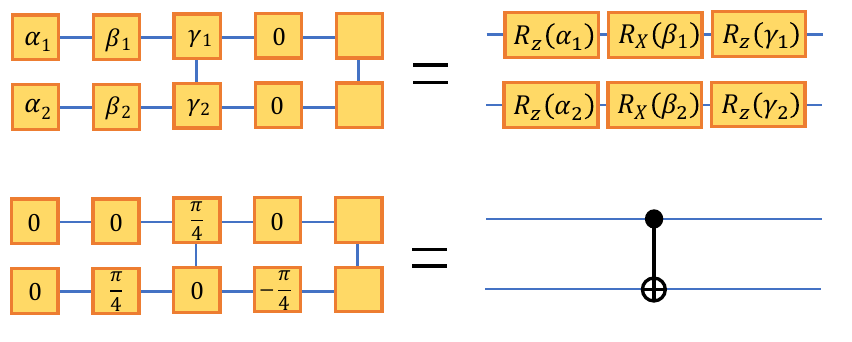}
\caption{The construction of the variational quantum classifier in our work including the construction of single-qubit gates (up) and controlled-NOT gates (down) adapted from \cite{Broadbent2009Universal}, where the corrections that consider the measurement outcomes are not explicitly shown.}
\label{UBQC_2}
\end{figure}

Since this protocol is universal for quantum computation,
we can construct a variety of circuits for different uses.
In our work,
we take the advantage that it is efficient to construct single-qubit gates and controlled-NOT gates in this framework as illustrated in Fig.~\ref{UBQC_2} to build a variational quantum classifier as already shown in Fig.~\ref{VQC}.
In Fig.~\ref{UBQC_2},
the angles on the left side denotes the measurement angles to do the measurement based computation,
where the corrections that consider the measurement outcomes are not explicitly shown.
Then the achieved single-qubit gates and controlled-NOT gates are presented on the right side.
In this setting,
the parameters encoded in the single-qubit rotations are continuous and the controlled-NOT gates are fixed.
So in order to encrypt the continuous information,
the $\theta_{x, y}$ should be sampled from a continuous set such as $R[0,2\pi]$ we adopt.

\section{Details of our numerical experiments}

\emph{Data encoding.}\textemdash
In the main text,
the data of handwritten digits and breast cancer are encoded as the input of the circuit with the method of amplitude encoding,
which enables us to represent $N=2^n$ dimensional classical data with $n$ qubits.
During the private delegated computing process,
Alice can only prepare single qubit states and send them to Bob.
So it points out the need to construct highly-entangled target states on Bob's side.
According to the work from 
\cite{Bergholm2005Quantum,Plesch2011Quantum},
we can decompose any quantum state into a series of single-qubit rotations and two-qubit gates like controlled-NOT gates.
The single-qubit rotations can be directly created in the structure of UBQC. As for the controlled-NOT gates, 
if the two qubits are adjacent, 
the controlled-NOT gates can be directly constructed as shown before. 
Otherwise if the two qubits are not adjacent, 
we can use SWAP gates and controlled-NOT gates to accomplish the target controlled-NOT gate where the SWAP gates can also be constructed using three controlled-NOT gates.
In this way,
we can use the protocol above as a subroutine to achieve the amplitude encoding strategy.

\emph{Gradient descent.}\textemdash
In the optimization process,
we set the loss function $L$ as cross entropy which takes the expectation value of the ancillary qubits in the standard basis as a component.
The output state of the ancillary qubits can be written as $\rho_{out}$.
Let $\mathbf{g} \equiv\left(g_{1}, \ldots, g_{2^{m}}\right)=\operatorname{diag}\left(\rho_{\mathrm{out}}\right)$ and $\mathbf{a}$ be the label corresponding to the input state $\ket{\phi}_{\mathbf{in}}$ in the form of one-hot encoding.
Then we can formalize the loss function as
$$
L\left(h\left(|\psi\rangle_{\mathrm{in}} ; \Theta\right), \mathbf{a}\right)=-\sum_{k} a_{k} \log g_{k}.
$$
It can be seen that reducing the loss function is consistent with reducing the distance between $\mathbf{a}$ and $\mathbf{g}$.
To reduce the loss function,
computing the derivatives of $L$ with respect to the circuit parameters can be transformed into computing the derivatives of some expectation values with respect to these circuit parameters according to the chain rule.
That is,
$$
\frac{\partial L\left(h\left(|\psi\rangle_{\mathrm{in}} ; \Theta\right), \mathbf{a}\right)}{\partial \theta} = -\sum_{k} \frac{a_{k}}{g_{k}} \frac{\partial g_{k}}{\partial \theta}.
$$
The next step that computes the derivatives of $g_k$ to the circuit parameters can be done with the "parameter shift rule" since $g_k$ can be seen as an expectation value of an observable which we denote as $B_k$ here
\cite{Mitarai2018Quantum,Li2017Hybrid,Schuld2019Evaluating}:
$$
\frac{\partial g_k}{\partial \theta}=\frac{\partial \langle B_k\rangle}{\partial \theta}=\frac{\langle B_k\rangle^{+}-\langle B_k\rangle^{-}}{2},
$$
where $\langle B_k\rangle^{\pm}$ denotes the expectation value of $B_k$ with the parameter $\theta$ being $\theta \pm \frac{\pi}{2}$.
With these preparations,
we can optimize the variational quantum circuit with gradient based methods.

\emph{Numerical settings.}\textemdash
In the main text,
we present the numerical simulations for single-party delegated learning and multiparty distributed learning with two datasets.
Here in Table \ref{T0} and Table \ref{T1} we list some numerical settings for these experiments.
 Moreover, the numerical simulations are carried out using the Yao.jl framework \cite{Luo2020Yao}.

\begin{table}[htpb]
\begin{tabular}{|l|l|l|}
\hline
                    & mnist classification & tumour diagnosis \\ \hline
classes          & 2("1" or "9")          & 2              \\ \hline                   
circuit depth    & 30                   & 30              \\ \hline
num of qubits     & 8                    & 6                \\ \hline
learning rate     & 0.001                & 0.001           \\ \hline
num of iterations & 300                  & 300             \\ \hline
batch size        & 50                  & 100              \\ \hline
optimizer        & Adam                 & Adam             \\ \hline
\end{tabular}
\caption{Parameter settings for single-party delegated learning.}
\label{T0}
\end{table}

\begin{table}[htbp]
\centering
\begin{tabular}{|l|l|l|}
\hline
                          & mnist classification  & tumour diagnosis\\ \hline
classes                  & 2("1" or "9")          & 2              \\ \hline 
gradient bound           & 0.01                  & 0.01            \\ \hline
circuit depth            & 30                    & 30              \\ \hline
num of qubits             & 8                     & 6               \\ \hline
learning rate             & 0.001                 & 0.001           \\ \hline
num of iterations         & 1500                  & 1000            \\ \hline
num of clients            & 10                    & 4               \\ \hline
training set(per client)  & 200                   & 100             \\ \hline
batch size(per client)    & 50                    & 50              \\ \hline
optimizer                 & Adam                  & Adam            \\ \hline
\end{tabular}
\caption{Parameter settings for multiparty distributed learning.}
\label{T1}
\end{table}

\section{Robustness test for the variational quantum classifiers}

Considering the inevitable noise in the real quantum devices,
both the qubits sent by Alice and the computation executed by Bob may deviate from the noiseless situation.
To explore how well the classifier we have proposed performs against the random noise,
we carry out several numerical simulations with noise included.

\begin{figure}
\includegraphics[width=0.48\textwidth]{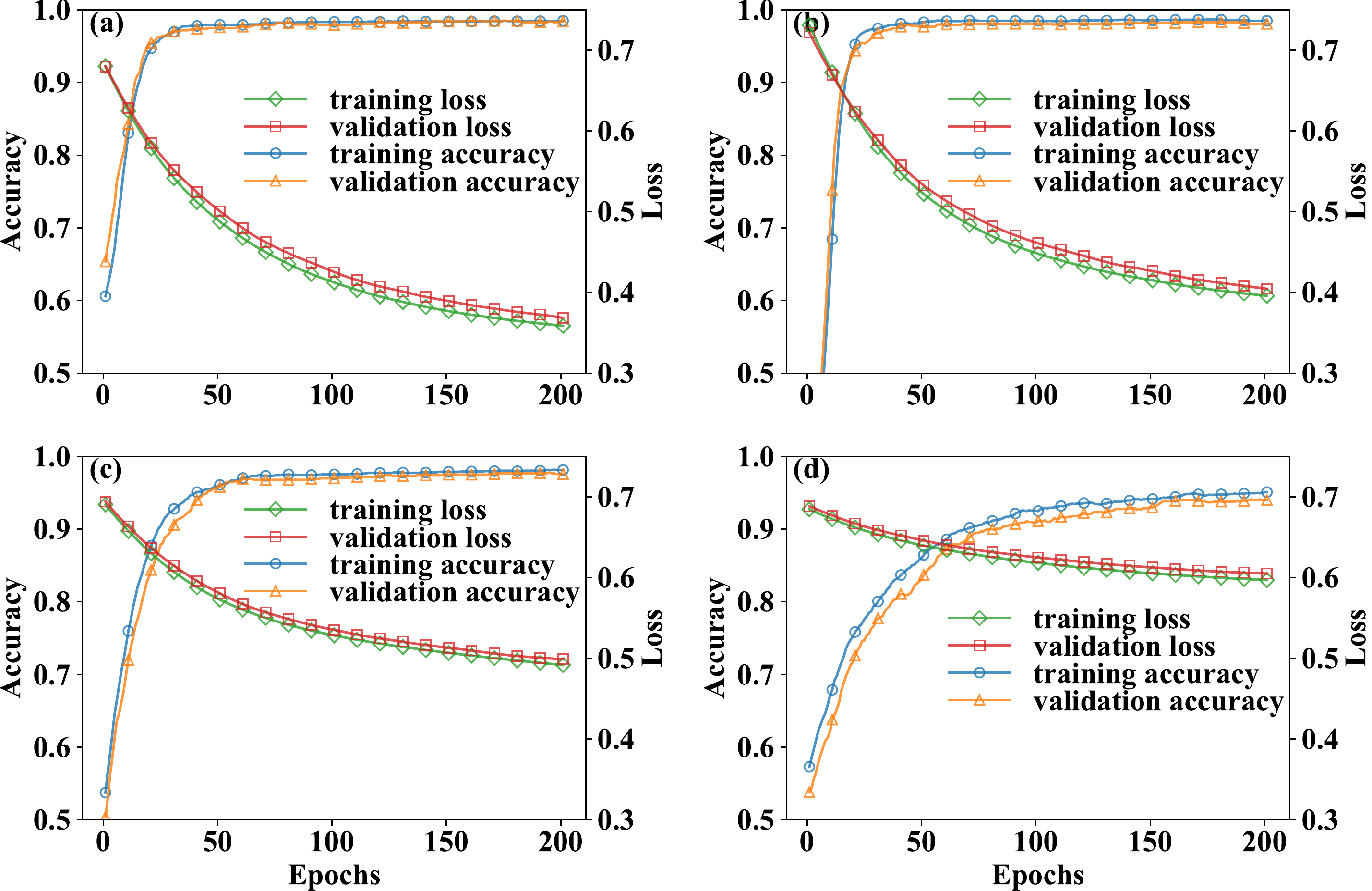}
\caption{The robustness test with the handwritten digits from the MNIST dataset: (a) The zero noise situation corresponding to Fig.~\ref{VQC_train}(a). (b) The noise strength of the input state is set as $0.02$. (c) The noise strength of the input state is set as $0.05$. (d) The noise strength of the input state is set as $0.1$.}
\label{VQC_robustness_mnist}
\end{figure}

\begin{figure}
\includegraphics[width=0.48\textwidth]{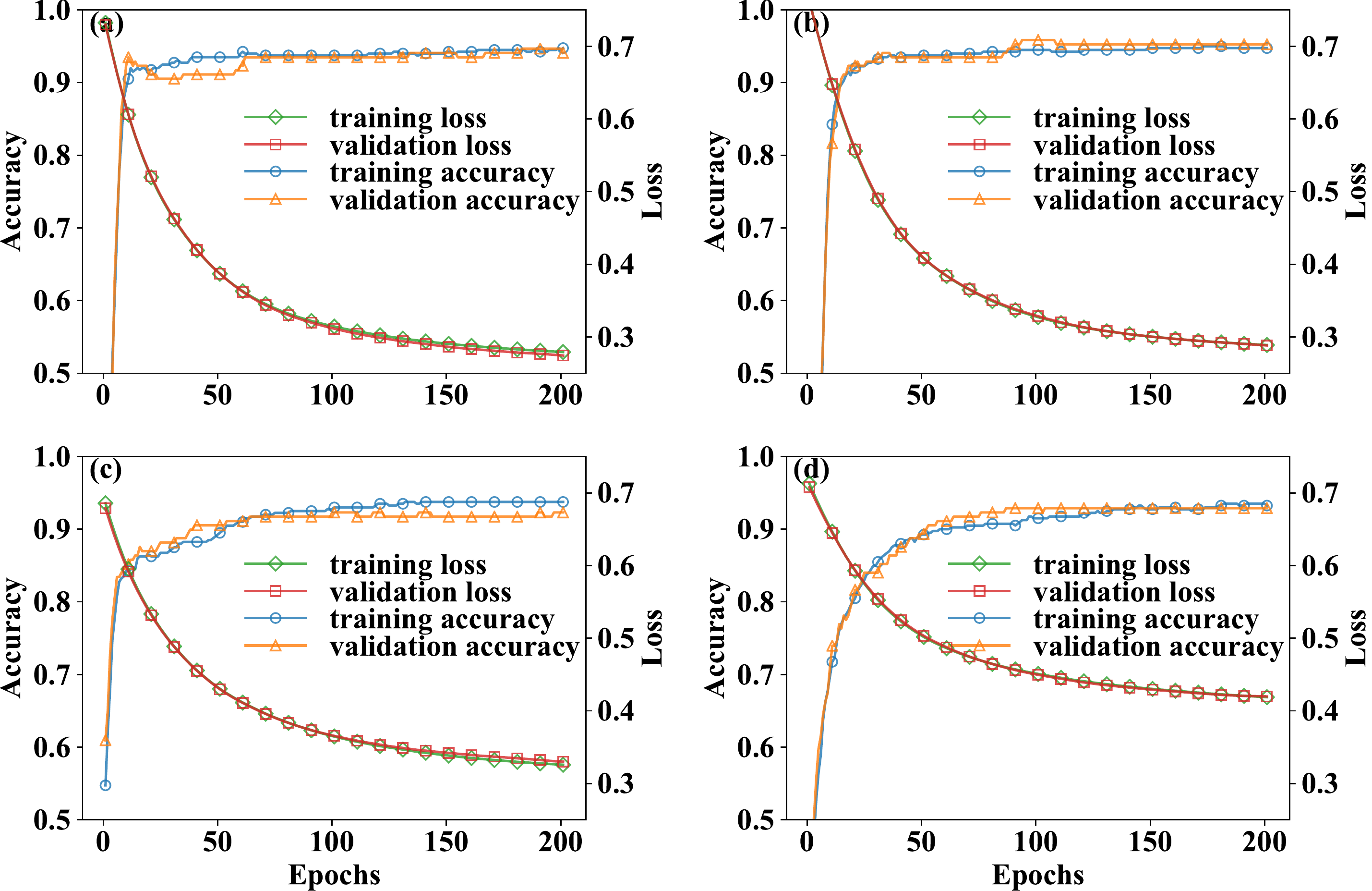}
\caption{The robustness test with the data from the Wisconsin Diagnostic Breast Cancer dataset: (a) The zero noise situation corresponding to Fig.~\ref{VQC_train}(b). (b) The noise strength of the input state is set as $0.02$. (c) The noise strength of the input state is set as $0.05$. (d) The noise strength of the input state is set as $0.1$.}
\label{VQC_robustness_cancer}
\end{figure}

We divide the noise into two parts: 
(1) the noise in the input training data, which may result from the qubits sent by Alice and the measurements to generate the entangled state by Bob; 
(2) the noise in the parameterized circuit, which may result from the qubits sent by Alice and the measurements to do the computation by Bob. 
For the first part, 
we add noise, 
which is a number normally distributed with mean value zero and standard deviation one times the input noise strength that we can set with different values, 
to every element of the input vector.
For the second part,
we add noise, 
which is $0.001$ times a number normally distributed with mean value zero and standard deviation one, 
to every element of the gradient independently during the training process.
The training process for two datasets are shown in Fig.~\ref{VQC_robustness_mnist} and Fig.~\ref{VQC_robustness_cancer},
where the model can still achieve a decent performance when the noise strength is not too large.
For concreteness,
in both the MNIST dataset and the Wisconsin Diagnostic Breast Cancer dataset,
the variational quantum learning model shows good resilience to noise when the noise strength is below $0.02$.
With the increase of the noise,
the accuracy will drop and it will take more time to converge.

\section{Differential privacy with different noise strengths}

\begin{figure}
\includegraphics[width=0.48\textwidth]{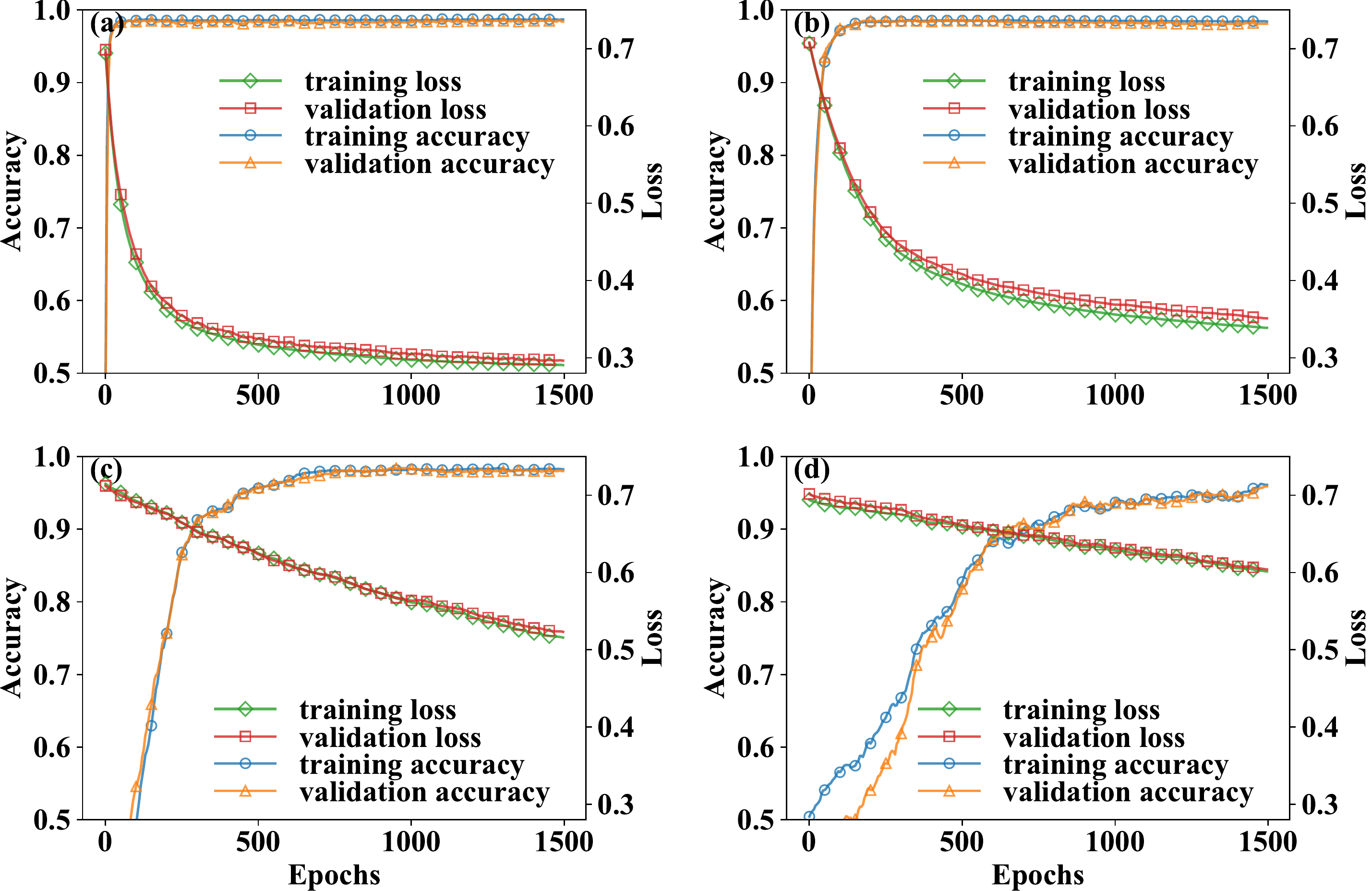}
\caption{The performance of our quantum federated learning protocol with the adaption of differential privacy for classifying the handwritten digits from the MNIST dataset: (a) The zero noise situation $\mu = + \infty$. (b) The noise strength $\mu = 1$. (c) The noise strength $\mu = 0.1$. (d) The noise strength $\mu = 0.05$. There is more noise when $\mu$ is smaller as defined in Algorithm \ref{Algo_PDL}.}
\label{VQC_dp_mnist}
\end{figure}

\begin{figure}
\includegraphics[width=0.48\textwidth]{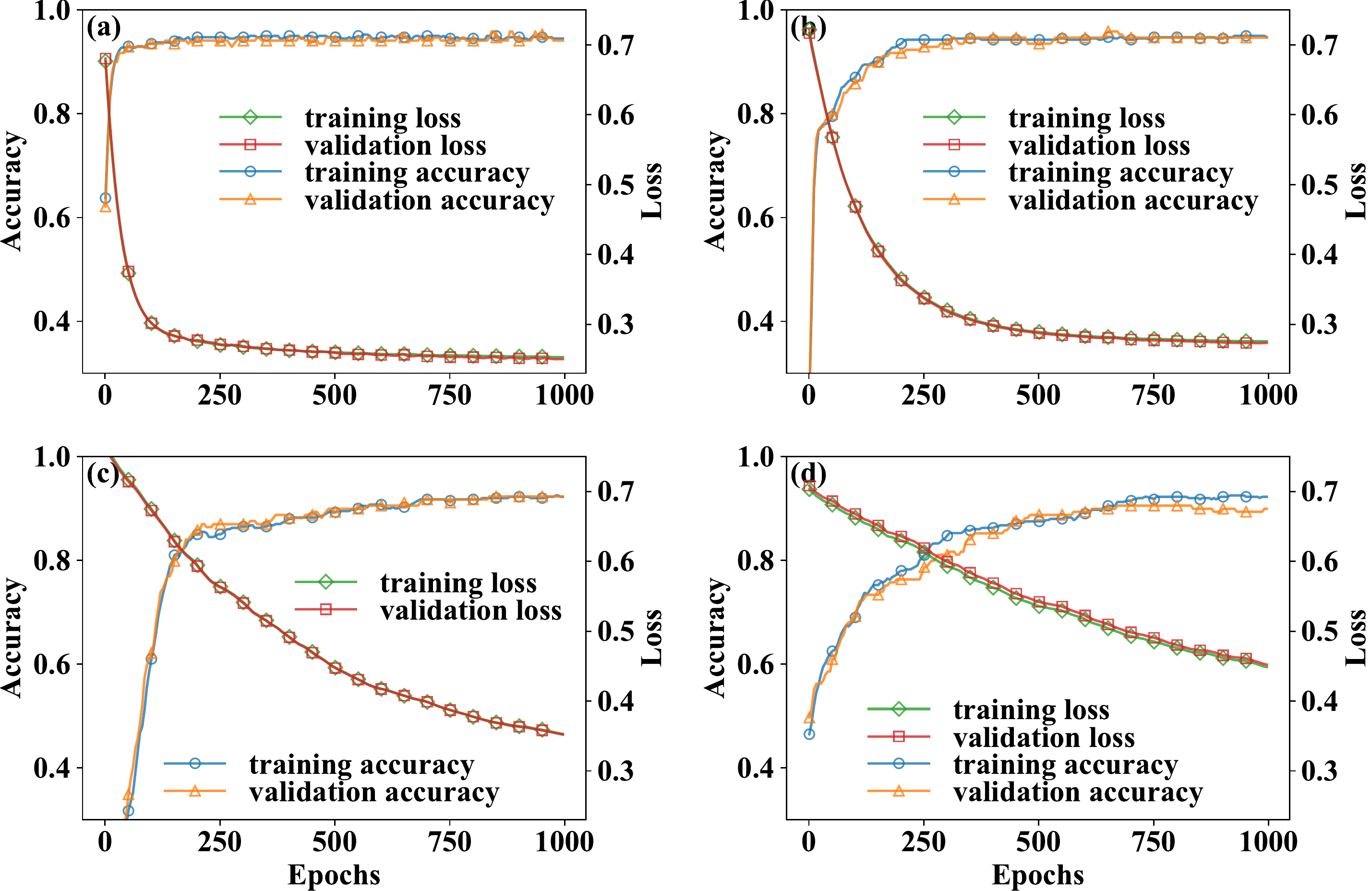}
\caption{The performance of our quantum federated learning protocol with the adaption of differential privacy for classifying the data from the Wisconsin Diagnostic Breast Cancer dataset: (a) The zero noise situation $\mu = + \infty$. (b) The noise strength $\mu = 1$. (c) The noise strength $\mu = 0.2$. (d) The noise strength $\mu = 0.1$. There is more noise when $\mu$ is smaller as defined in Algorithm \ref{Algo_PDL}.}
\label{VQC_dp_cancer}
\end{figure}

The noise in the differential privacy plays a beneficial role to protect the privacy,
which is different from the noise in the robustness test section which may harm the training.
The work in \cite{Sheng2018Blind} has considered the feasibility of blind quantum computation in a noise channel, and the work in \cite{Chen2021Federated} has explored the potential of noise in the federated quantum machine learning.
In this section,
we provide the numerical simulations for the differential privacy scheme with different noise strengths.
The basic settings have already been given in Table \ref{T1}.
For the noise strength $\mu$ defined in Algorithm \ref{Algo_PDL}, 
we also set four different values,
where $\mu = + \infty$ means no noise.
The accuracy and loss as a function of epochs during the training process for the two datasets are shown in Fig.~\ref{VQC_dp_mnist} and Fig.~\ref{VQC_dp_cancer}.

\begin{figure}
\includegraphics[width=0.48\textwidth]{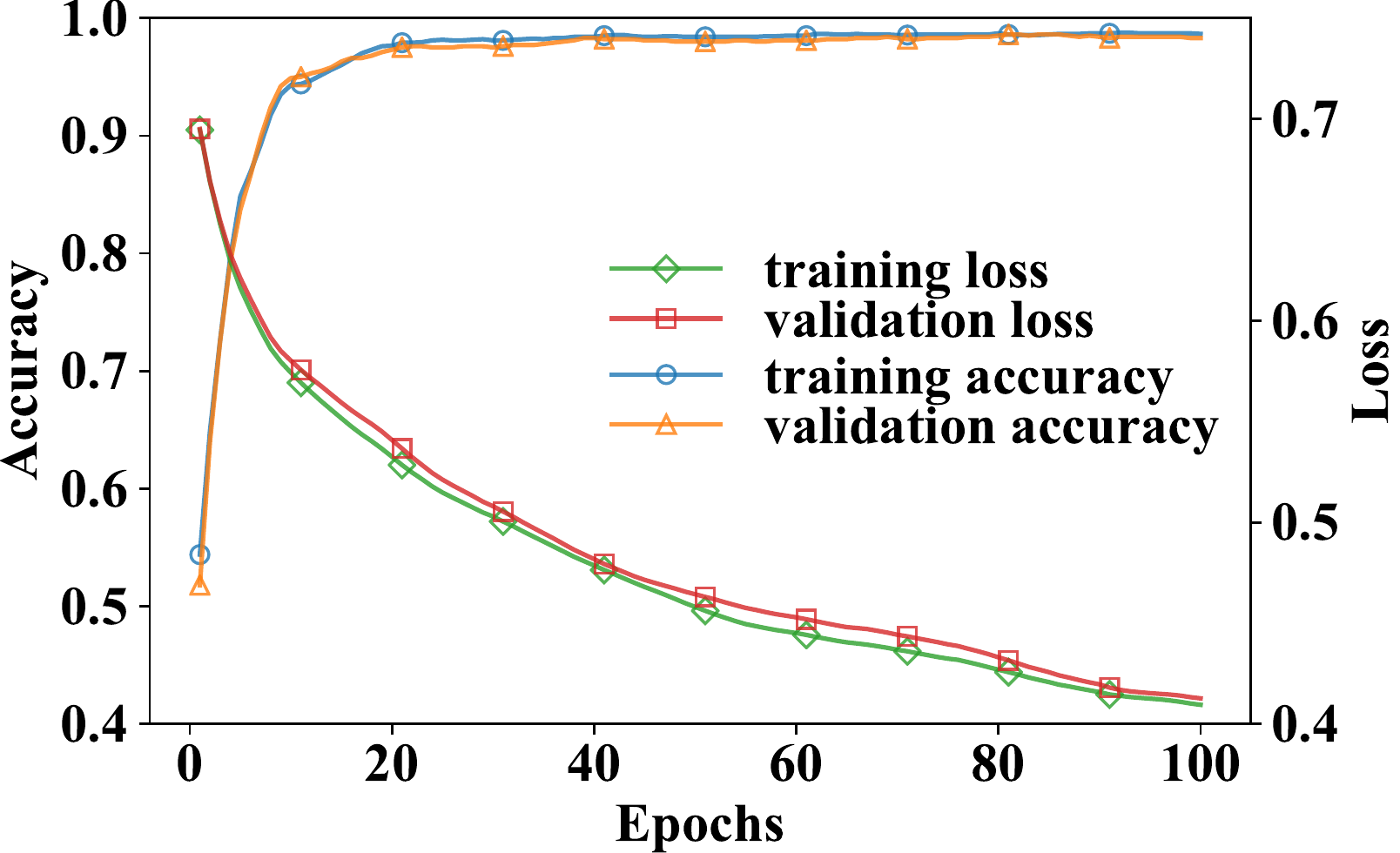}
\caption{The performance of our protocol under the situation that the prepared input states from the different parties have different noise levels. We use the differential privacy parameter $\mu$ to characterize the strength of the total noises. The noise levels for ten different clients are set as $\mu$ = [1,0.95,0.9,0.85,0.8,0.75,0.7,0.65,0.6,0.55], respectively.}
\label{VQC_dp_mnist_revision}
\end{figure}

It is apparent from the training curves that when we add more noise, 
the training accuracy will decrease and it will take more steps to converge for both the two datasets.
When the noise strength $\mu$ is $0.05$ in the MNIST dataset and $0.1$ in the Wisconsin Diagnostic Breast Cancer dataset,
the training becomes very slow and has the trend to fail with increasing noise.
Thus the results agree with the point that it is a trade-off between accuracy and privacy.

Furthermore, considering the real-life conditions, the different parties may have different quantum capabilities. So the intrinsic noises from the different parties may be different.
In order to test the robustness of the protocol in this situation,
we assign different noise strengths $\mu$ to different parties and assume that $\mu$ represents the combination of the intrinsic quantum noises and the artificially added noises.
We set the learning rate as $0.01$ and the rest settings are the same as that in Table~\ref{T1}.
We carry out  numerical simulations as shown in Fig.~\ref{VQC_dp_mnist_revision}. For this figure, it is evident that our protocol is still robust under this situation. 

\section{Encoding strategy with rotation angles}

\begin{figure}
\includegraphics[width=0.48\textwidth]{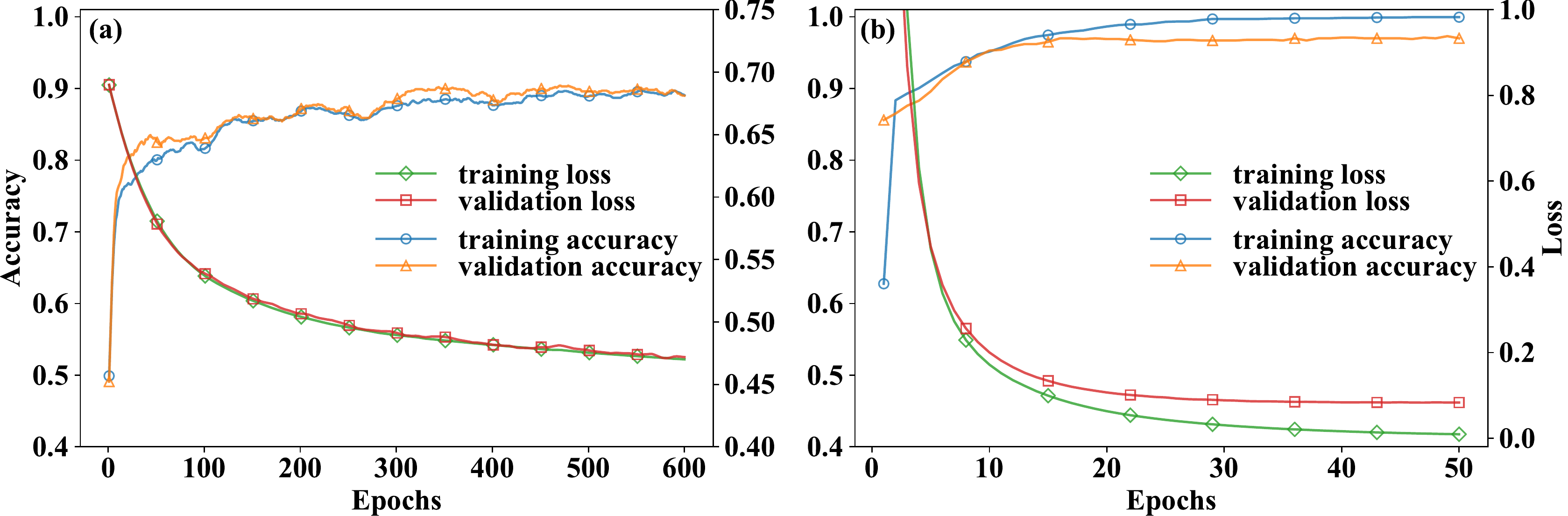}
\caption{(a) The training process for the data from MNIST dataset with the encoding strategy that encodes the input features into the circuit parameters. (b) The training process of classical neural networks with two hidden layers to handle the same data as that in (a), which proves that the training data from the circuit encoding process is separable.}
\label{VQC_R_encode}
\end{figure}

In the numerical simulations above,
we take the amplitude encoding strategy to encode the input features.
Although the client needs to create these entangled states with methods such as the one from \cite{Plesch2011Quantum},
it is not efficient compared with the strategy that encodes the features into the circuit parameters.
In this section,
we will give some brief simulations of the variational quantum classifiers with a different encoding strategy.

Firstly,
we still use the parameterized circuit shown in Fig.~\ref{VQC} to encode the input features.
The circuit starts with a zero state $\ket{00...0}$ as the input.
For the encoding process,
we use the framework from Fig.~\ref{VQC} with qubit number $8$ and depth $11$ which has $272$ parameters and initially set these parameters to be $0$.
Then for the $256$-dimensional features from the MNIST dataset,
we encode them into the first $256$ parameters among the $272$ parameters.
Then the outputs of the encoding circuit which is a set of $256$-dimensional vectors can be used as the inputs for the training circuit and the training can be carried out.
To better compare the performance with the amplitude encoding,
we take the same settings as the ones from Table \ref{T0},
and the training curves are given in Fig.~\ref{VQC_R_encode}(a).
It can be seen that the trained model can only achieve an accuracy of about $90\%$,
which falls behind the performance with the amplitude encoding strategy.
We further demonstrate that,
if we extract the outputs of the encoding circuit to feed in a 2-layer classical neural network,
the classical model can provide a decent performance as shown in Fig.~\ref{VQC_R_encode}(b).

We use this example to demonstrate that,
there are different ways to encode the input features for the training.
However,
on the one hand,
the amplitude encoding can achieve better performance but it requires the inefficient process to prepare the highly entangled states in the delegated tasks.
On the other hand,
the way we encode the features into the circuit parameters is more efficient but may suffer from the accuracy loss,
which calls for better encoding strategies for the future applications.
Furthermore,
the comparison of the performance of the variational model with the classical model indicates that we need to develop variational quantum learning models with higher expressive power to handle more complex tasks.

\end{document}